\documentclass[aps,preprint,showpacs,superscriptaddress,groupedaddress]{revtex4}  % for double-spaced preprint
\usepackage{graphicx}  % needed for figures
\usepackage{dcolumn}   % needed for some tables
\usepackage{bm}        % for math
\usepackage{amssymb}   % for math

\usepackage{color}
\usepackage{amsmath}

\usepackage{epsfig}
\usepackage{tikz}
\usetikzlibrary{decorations.pathmorphing}
 \usepackage{epstopdf} %%package to overcome problem with eps in pdf files
\newcommand{\hemvec}[1]{\mathbf{#1}}    %Bold vectors
\newcommand{\myint}[2]{\int \mathrm{d}#2\, \, #1 }

\newcommand{\fref}[1]{Figure~\ref{#1}}
\newcommand{\tref}[1]{Table~\ref{#1}}
\newcommand{\eref}[1]{Equation~\ref{#1}}
\newcommand{\cref}[1]{Chapter~\ref{#1}}

\newcommand{\erefs}[2]{Equations~\ref{#1} and~\ref{#2}}
\newcommand{\erefsto}[2]{Equations~\ref{#1}-\ref{#2}}

\begin{document}

\title{Estimation of correlation energy for excited-states of atoms}
\author{M.~Hemanadhan}
\affiliation{Department of Physics, Indian Institute of Technology, Kanpur 208016, INDIA}
\author{Manoj~K.~Harbola}
\affiliation{Department of Physics, Indian Institute of Technology, Kanpur 208016, INDIA}
\date{\today}
\begin{abstract}
The correlation energies of various atoms in their excited-states are estimated 
by modelling the Coulomb hole following the previous 
work by Chakravorty and Clementi. 
The parameter in the model is fixed by making 
the corresponding Coulomb hole to satisfy the 
exact constraint of charge neutrality. 
%The correlation energies obtained ...
\end{abstract}

\maketitle

\section{Correlation energy}
Electron correlation in many-electron system is of two kinds, 
%%one due to Coulomb interaction and the other due to 
%%exchange interactions.
%The quantum-mechanical nature of electron-electron interaction 
%is usually classified into two categories: 
one due to the Coulombic repulsion between the electrons and the other 
due to Fermi-Dirac statistics of electrons --
referred as Coulomb and Pauli correlations, respectively. 
%The classical Coulombic repulsion between electrons leads to 
%the lowering of the quantum mechanical probability of electrons approaching each other --
%However, 
%%Both of these correlations %repulsive terms 
%%lower the  quantum mechanical probability of electrons approaching each other -- 
%%creating a hole in the probability density around each electron -- 
%%the former is known as Coulomb hole and the latter Fermi hole.
%, avoiding two electrons with the same spin 
%approaching each other. 
%Pauli interaction (Fermi hole), and the Coulomb  can b
Coulomb correlations cannot be treated exactly 
as the precise form of the wavefunction 
for a many-electron system cannot be determined 
since the Schr\"{o}dinger equation for a many-electron system 
is not solvable. On the other hand, 
the effects of Pauli correlation can be explicitly taken 
care of by ensuring the wavefunction to be antisymmetric with respect 
to the interexchange of electron coordinates. 
For example, 
%\subsection{From Hartree-Fock model and beyond it}
in the Hartree-Fock treatment of the many-electron problem, 
the wavefunction is made antisymmetric by writing it as a 
Slater determinant in terms of single-particle orbitals. 
%the Fermi hole is taken care 
%the mutual repulsion between two parallel spins is partially 
%treated 
%by enforcing the wavefunction to be anti-symmetric 
%achieved by single Slater determinant. 
%However, the Coulomb hole is ignored. 
% mutual repulsion between electrons with antiparallel spins 
%is not fully taken into account. 
%%The left over 
%correlation energy is 
%The 
%%quantum chemical correlation energy, 
%%traditionally referred as the correlation energy $E^{QC}_c$, 
%%is therefore given 
%is traditionally given 
%defined 
The difference between 
the exact non-relativistic energy $E_{exact}^{NR}$ 
(which may be calculated to high accuracy by various techniques) 
and the Hartree-Fock energy $E_{HF}$
is traditionally referred as the correlation energy $E^{QC}_c$, 
and is given as 
\begin{align}
	E^{QC}_c	= E_{exact}^{NR} - E_{HF}.
%	E_c	= E_{\text{exact}}^{\text{NR}} - E_{\text{HF}}
	\label{eq:ecorr-def}
\end{align}
$E^{QC}_c$ will always be negative because the Hartree-Fock energy 
is an upper bound to the exact energy by the variational principle. 
Although the correlation energy is small compared to the total 
energy, its inclusion is important as in the  
ionization potential, electron affinities, excitation energy calculations. 
Obtaining $E_c$ is one of the challenges in many-electron problem.
%Approximating the $E_c$ is a challenging task. 
In the following sections, we present some of our attempts 
to estimate the correlation energies of atoms in 
ground- and excited-states. 

\subsection{Lee-Yang-Parr (LYP) correlation energy functional}
%\subsection{Colle-Salvetti formula, and Lee-Yang-Parr (LYP) correlation}
%\subsection{Based on LYP}
%A correlation energy formulae due to Colle Salvetti, 
%in which the correlation energy density is expressed in terms of the electron 
%density and a Laplacian of the second order HF 
%density matrix, is restated as a formula involving the 
%density and the local kinetic energy density.
A correlation energy formulae due to Colle Salvetti (CS)~\cite{colle-salvetti:1975}, 
%colle-montagnani-riani-salvetti:1978
in which the correlation energy density is obtained from an approximate 
correlated wavefunction,  
%Colle Salvetti formula for correlation energy 
%and its adaptation 
was adapted to density functional form by 
Lee, Yang and Parr (LYP)~\cite{lee-yang-parr:1988}, 
and is given for ground-states by the formula 
%The LYP formula for correlation energy of atoms in ground state is given by,
\begin{align}
	E_c^{\text{LYP}}
	= -a 
	\myint{
		\frac{
			\rho(\hemvec{r})
			+ 2b\rho(\hemvec{r})^{-5/3}
			\left[
			\rho_{\alpha}(\hemvec{r}) t^{\alpha}_{HF}
			+ \rho_{\beta}(\hemvec{r}) t^{\beta}_{HF}
			- \rho(\hemvec{r}) t_w(\hemvec{r})
			\right]
			e^{-c\rho(\hemvec{r})^{1/3}}
		}{
			1 + d\rho(\hemvec{r})^{-1/2}
		} 
		\gamma(\hemvec{r})
	}{
	\hemvec{r}
	}
	\label{eq:ec-lyp}
\end{align}
where parameters $a,b,c,$ and $d$ are chosen to get 
the correlation energy of the ground-state of He atom, and 
\begin{align}
	\gamma(\hemvec{r})
	= 2
	\left[
		1 - 
		\frac{
			\rho_{\alpha}^2(\hemvec{r})
			+\rho_{\beta}^2(\hemvec{r})
		}{
		\rho^2(\hemvec{r})
		}
	\right]
\end{align}
is a dimensionless constant. 
The Hartree-Fock kinetic energy density corresponding to 
%Hartree-Fock kinetic energy density for 
up-spin electron ($t^{\alpha}$) is given by 
\begin{align}
	t^{\alpha}(\hemvec{r}) 
	= \frac{1}{2}
	t_{HF}(2\rho_{\alpha}(\hemvec{r}),\hemvec{r})
\end{align}
Similarly, the corresponding kinetic energy density ($t^{\beta}$) expression 
for the down-spin electron is 
%is the Hartree-Fock kinetic energy density for spin-up,
\begin{align}
%	&  & 
	t^{\beta}(\hemvec{r}) 
	= \frac{1}{2}
	t_{HF}(2\rho_{\beta}(\hemvec{r}),\hemvec{r})
\end{align}
The total Hartree-Fock kinetic energy density ($t_{HF}$) is given by 
\begin{align}
	t_{HF}
	= t_{TF}
	+ \frac{1}{9} t_W(\hemvec{r}) 
	+ \frac{1}{18}\nabla^2 \rho
\end{align}
where $t_{TF},$ and $t_{W}$ are the kinetic energy densities by Thomas-Fermi 
%corresponding to zeroth-(Thomas-Fermi) 
and Weizs\"{a}cker respectively, and is given by 
\begin{align}
	t_{TF}
	= \frac{3}{10}
	\left(3\pi^2\right)^{2/3}
	\rho^{5/3}	\\
%\end{align}
%is Thomas-Fermi Kinetic energy density. 
%\begin{align}
	t_{W}
	= \frac{1}{8}
	\frac{\left|\nabla \rho\right|^2}{\rho}
	-\frac{1}{8}
	\nabla^2 \rho
\end{align}
%is Weizsacker correction [25] to kinetic energy density.
It has been shown that the $E_c^{\text{LYP}}$ gives 
atomic correlation energies for ground-states within a few percent 
of their accurate values. 
%give reasonable good estimation of correlation energy for ground states 
%of atoms and molecules. 
%However, the LYP functional does not reduce to LSDA for uniform densities.
LYP functional has been employed to calculate energies of excited-states 
of atoms using Harbola-Sahni orbitals~\cite{harbola-sahni:1989a,roy-jalbout:2007}.

Attempts to estimate correlation energies for excited-states 
by extending the LYP functional using the method 
of splitting $k$-space was pursued recently~\cite{thesis:shamim}. 
This is based on the observation that 
the derivation of Colle-Salvetti and LYP formulae are quite general, 
and the ideas are equally 
%therefore the should be 
applicable to excited states also. 
The modified LYP functional for an excited state corresponding to 
one-gap system 
%~(\fref{fig:cosh-occ}) 
is obtained by replacing 
$t_{TF}$ and $t_{W}$ in~\eref{eq:ec-lyp} 
with the modified Thomas Fermi kinetic energy density ($t_{mTF}$)
%~(\eref{eq:mtf-cosh}) 
\begin{align}
	t_{mTF}
	= \frac{3}{10}
	\left(3\pi^2\right)^{2/3}
	\left[
		\rho_3^{5/3} - \rho_2^{5/3} + \rho_1^{5/3}
	\right]
\end{align}
% is given by,
%For excited states we use modified expressions
%for kinetic energy term 
and the modified Weizsacker term ($t_{mW}$)
%~(\eref{eq:mgea2-cosh})
%and the Weizsacker correction term
\begin{align}
	t_{mW}
	= 
	\frac{1}{8}
	\left[
	\frac{\left|\nabla \rho_1\right|^2}{\rho_1}
	+ 
	\frac{\left|\nabla \rho_3\right|^2}{\rho_3}
	- 
	\frac{\left|\nabla \rho_2\right|^2}{\rho_2}
	\right]
	-\frac{1}{8} 
	\left[
		\nabla^2 \rho_1
		+ \nabla^2 \rho_3
		- \nabla^2 \rho_2
	\right]
\end{align}
%are used in~\eref{eq:ec-lyp} 
%and expect to get reasonable correlation energy for excited states. 
%The aim of work in this appendix is to extend LYP functional 
%for correlation energy to excited states. 
%For this we follow the idea of splitting the 
%$k$-space in accordance with the occupation of 
%orbitals in excited-state.
The parameters ($a, b, c$ and $d$) in the modified LYP functional 
for the excited-state calculations 
are chosen to be same as in the ground-state calculations.  
%in the LYP functional~(\eref{eq:ec-lyp}) are chosen. 
%For the excited-state calculations using the modified LYP functional, 
% to get correct correlation energy for ground state of He atom. 
It is observed that the modified LYP functional 
%does not give desired results and 
leads to insignificant improvement over the 
correlation energy obtained with ground state
functional.
%Since the above mentioned approach does not seem to work, 
%we look for another simple method to achieve the goal. 
%Coll-Salvetti formulation for correlation energy has
%been obtained by choosing such values for parameters 
%Further,
In addition to chosing the ground-state parameters for the modified LYP functional, 
%Alternatively, the parameters in the modified LYP functional are 
a new set of parameters were also obtained by fitting 
%%Attempts have been made to 
%Therefore, these parameters might be different for excited states. 
%We did 
%%fit the parameters 
for a particular excited state of He. 
The correlation energies so obtained for the excited states of other atoms 
doesn't improve the results. 
%Again the method does not seem to work and 
This study indicates that some other approach should be adopted to 
%the choice of parameters plays a crucial step in 
estimate the correlation energies for excited states.
%in this procedure, 
%and is also state-dependent. 
%We conclude that for excited states, parameters $a, b, c$ and $d$ are state
%dependent therefore one particular values of these parameters may not work for all types

In the next section, we try to estimate the correlation energies following 
the previous work by Chakravorty and Clementi~\cite{chakravorty-clementi:1989}.
%The parameter in the model is fixed by making
%the corresponding Coulomb hole satisfy the
%exact constraint of charge neutrality.
%where the dependence of the parameter are minimized with the help of exact conditions.  

\subsection{Correlation energy by modelling pair-correlation function}
%\subsection{Based on Coulomb hole}
Chakravorty and Clementi~\cite{chakravorty-clementi:1989} 
proposed a method to include the Coulomb hole in the Hartree-Fock method. 
In this method, a soft-Coulomb hole of Gaussian nature is introduced in the 
expressions for Hartree-energy % ($E_H^{HF}$)
\begin{equation}
	E_{H}^{HF}
	= \frac{1}{2} \sum_{i,j} \iint 
		\frac{
		\psi^*_i(\mathbf{r})\psi_i(\mathbf{r}) \psi_j(\mathbf{r}')\psi^*_j(\mathbf{r}')
		}
		{\left|\mathbf{r}-\mathbf{r}'\right|} d\mathbf{r} d\mathbf{r}'	
	\label{eq:ecoul}
\end{equation}
and the 
exchange-energy %($E_x^{HF}$)
\begin{equation}
	E_{x}^{HF}
	= -\frac{1}{2} \sum_{i,j}{}^{'} \iint 
		\frac{
		\psi^*_{i,\sigma}(\mathbf{r})\psi^*_{j,\sigma}(\mathbf{r}') \psi_{i,\sigma}(\mathbf{r}')\psi_{j,\sigma}(\mathbf{r})
		}
		{\left|\mathbf{r}-\mathbf{r}'\right|} d\mathbf{r} d\mathbf{r}' .
	\label{eq:eexch}
\end{equation}
The modified expression for the corresponding energies are given by
%corresponding Hartree- and exchange-energy terms are given by 
\begin{align}
	E^{HF}_{H,\gamma}
	&= \frac{1}{2} \sum_{i,j} \iint 
		\frac{
		\psi^*_i(\mathbf{r})\psi_i(\mathbf{r}) \psi_j(\mathbf{r}')\psi^*_j(\mathbf{r}')
        	\left[1-\exp({-\gamma \left|\mathbf{r}-\mathbf{r}'\right|^2}) \right]
		}
		{\left|\mathbf{r}-\mathbf{r}'\right|} d\mathbf{r} d\mathbf{r}'	
	\label{eq:ecoul-clem}	\\
%\end{equation}
%and 
%\begin{equation}
	E^{HF}_{x,\gamma}
	&= -\frac{1}{2} \sum_{i,j}{}^{'} \iint 
		\frac{
		\psi^*_{i,\sigma}(\mathbf{r})\psi^*_{j,\sigma}(\mathbf{r}') \psi_{i,\sigma}(\mathbf{r}')\psi_{j,\sigma}(\mathbf{r})
        	\left[1-\exp({-\gamma \left|\mathbf{r}-\mathbf{r}'\right|^2}) \right]
		}
		{\left|\mathbf{r}-\mathbf{r}'\right|} d\mathbf{r} d\mathbf{r}'
	\label{eq:eexch-clem}
\end{align}
The parameter $\gamma$ determines the size of the Coulomb hole 
and is parameterized in their work~\cite{chakravorty-clementi:1989}. 
The above equation reduces to 
Hartree energy $E_H^{HF}$ and exchange-energy $E_x^{HF}$ 
of the Hartree-Fock model in the limit $\gamma=\infty$. 
The correlation energy is then obtained by 
        \begin{align}
        E_{c} = 
        (E^{HF}_{\textrm{H}} + E^{HF}_x)
        - 
%        \left. 
        (E_{\textrm{H},\gamma} + E_{x,\gamma})
%	\right|_{\text{Coulomb hole}} 
	\label{eq:ecorrhf}
        \end{align}
%However, one should still fix the parameter $\gamma$, and this is done by matching 
%$E_c$ with experimental values. 
%, and calculating $E_c$.
%among which are configuration interaction (CI), many-body perturbation theory 
%and density functional method. 
%Hartree-Fock treatment of the many-electron problem, 
%The density functional methods provides an solution for this ... and tremendeous attempts 
%have been made to approximate the corresponsing correlation functional. 
%\subsection{From density-functional framework}
Like in traditional quantum theory, in the density-functional theory too, the exact exchange-correlation energy functional 
can be mathematically expressed as
        \begin{align}
        E_{xc}[\rho] 
	= \frac{1}{2}
        \iint 
        \frac{\rho(\mathbf{r}_1) 
        \rho_{xc}(\mathbf{r}_1,\mathbf{r}_2)}{\left|\mathbf{r}_1-\mathbf{r}_2 \right|}  d\mathbf{r}_1 d\mathbf{r}_2
        \end{align}
where,
$\rho_{xc}(\mathbf{r}_1,\mathbf{r}_2)$ is the exchange-correlation hole. 
The difference in the traditional correlation energies and the DFT correlation energies are numerically very small. 
The exchange- and correlation- holes are usually decoupled as 
$\rho_{xc}(\mathbf{r}_1,\mathbf{r}_2) = \rho_{x}(\mathbf{r}_1,\mathbf{r}_2) + \rho_{c}(\mathbf{r}_1,\mathbf{r}_2)$. 
In terms of exchange-hole, the exchange-energy functional is given by 
\begin{align}
	E^{DFT}_x[\rho]	
	&= \frac{1}{2}
        \iint 
        \frac{\rho(\mathbf{r}_1) 
        \rho_{x}(\mathbf{r}_1,\mathbf{r}_2)}{\left|\mathbf{r}_1-\mathbf{r}_2 \right|}  d\mathbf{r}_1 d\mathbf{r}_2
\end{align}
and the corresponding correlation-energy functional in terms of correlation-hole is 
%the correlation energy functional is given by 
\begin{align}
	E^{DFT}_c[\rho]	
%	&= E^{\lambda=1}[\rho] - E^{\lambda=1}_{HF}[\rho]	\\
	&= \frac{1}{2}
        \iint 
        \frac{\rho(\mathbf{r}_1) 
        \rho_{c}(\mathbf{r}_1,\mathbf{r}_2)}{\left|\mathbf{r}_1-\mathbf{r}_2 \right|}  d\mathbf{r}_1 d\mathbf{r}_2
	\label{eq:ecdft}
\end{align}
%where $\lambda$ is the interelectronic interaction strength. 
%The difference between $E^{QC}_c[\rho]$ and $E^{DFT}_c[\rho]$ is very small numerically. 
The explicit dependence of Coulomb correlation hole $\rho_{c}(\mathbf{r}_1,\mathbf{r}_2)$ 
on density $\rho$ is unknown and has to be approximated. 
However, the constraints to be satisfied by the $\rho_{c}(\mathbf{r}_1,\mathbf{r}_2)$ are known and 
are obtained from the exact constraints on the $\rho_{xc}(\mathbf{r}_1,\mathbf{r}_2)$ and $\rho_{x}(\mathbf{r}_1,\mathbf{r}_2)$: 
	\begin{subequations}
        \begin{align}
        \lim_{r_{12} \rightarrow \infty} \frac{\rho_{xc}(\mathbf{r}_1,\mathbf{r}_2)}{\rho(\mathbf{r}_2)} &= 0	
%	\\
	&
        \lim_{r_{12} \rightarrow \infty} \frac{\rho_{x}(\mathbf{r}_1,\mathbf{r}_2)}{\rho(\mathbf{r}_2)}	&= 0 
	\\	
        \lim_{r_{12} \rightarrow 0} \frac{\rho_{xc}(\mathbf{r}_1,\mathbf{r}_2)}{\rho(\mathbf{r}_2)} &= -1 
%	\\
	&
        \lim_{r_{12} \rightarrow 0} \frac{\rho_{x}(\mathbf{r}_1,\mathbf{r}_2)}{\rho(\mathbf{r}_2)} &= -\frac{1}{2}
	\\
        \int \rho_{xc}(\mathbf{r}_1,\mathbf{r}_2) d\mathbf{r}_2 &=-1 
	& 
	\int \rho_x(\mathbf{r}_1,\mathbf{r}_2) d\mathbf{r}_2 &=-1
        \end{align}
	\end{subequations}
These give the constraints on Coulomb hole $\rho_{c}(\mathbf{r}_1,\mathbf{r}_2)$ from 
$\rho_{c}(\mathbf{r}_1,\mathbf{r}_2)=\rho_{xc}(\mathbf{r}_1,\mathbf{r}_2)-\rho_{c}(\mathbf{r}_1,\mathbf{r}_2)$ as  
	\begin{subequations}
        \begin{align}
        \lim_{r_{12} \rightarrow \infty} \frac{\rho_{c}(\mathbf{r}_1,\mathbf{r}_2)}{\rho(\mathbf{r}_2)} &= 0 
	\label{eq:rhocorr-1}\\
        \lim_{r_{12} \rightarrow 0} \frac{\rho_{c}(\mathbf{r}_1,\mathbf{r}_2)}{\rho(\mathbf{r}_2)} &= -\frac{1}{2}
	\label{eq:rhocorr-2}\\
        \int \rho_c(\mathbf{r}_1,\mathbf{r}_2) d\mathbf{r}_2 &=0
	\label{eq:rhocorr-3}
        \end{align}
	\end{subequations}
%Thus approximate $\rho_{c}(\mathbf{r}_1,\mathbf{r}_2)$ has to satisfy these constraints. 
%Explicit dependence of Coulomb correlation hole on density is unknown and have to be approximated.
From~\erefsto{eq:ecoul}{eq:ecorrhf}, it is easily seen that the 
Coulomb hole $\rho_{c}(\mathbf{r}_1,\mathbf{r}_2)$ in the Chakravorty and Clementi method is  
        \begin{align}
        \rho_c(\mathbf{r}_1,\mathbf{r}_2) =
        \rho_c(\gamma,r_{12}) = 
        \left[
        - \rho(\mathbf{r}_2) + \rho_x(\mathbf{r}_1,\mathbf{r}_2)
        \right]
\exp({-\gamma \left|\mathbf{r}_1-\mathbf{r}_2\right|^2}) 
	\label{eq:coulombhole-chakra}
        \end{align}
where $r_{12}=\mathbf{r}_1-\mathbf{r}_2$. 
It is observed that the Coulomb hole in the Chakravorty and Clementi method does not satisfy the 
charge neutrality condition (\eref{eq:rhocorr-3}).  

In the next section, we try to model the correlation hole using the Yukawa form for the Coulomb hole 
along the same lines as the works by Chakravorty and Clementi. 
We, however, also put in additional terms to satisfy the charge neutrality condition. 

\section{Yukawa model for the Coulomb correlation hole}
The Hartree- ($E_{\textrm{H}}^{\text{Yuk},\gamma}$) and 
the exchange-energy ($E_{\textrm{H}}^{\text{Yuk},\gamma}$) 
obtained using the Yukawa form 
instead of Gaussian form in~\erefs{eq:ecoul-clem}{eq:eexch-clem} 
is given as 
%for the Coulomb hole are given by 
% following in the same lines as the works by Chakravorty and Clementi.
        \begin{equation}
        E_{\textrm{H}}^{\text{Yuk},\gamma}
        = \frac{1}{2} \iint 
                \frac{
        \rho(\mathbf{r}_1) \rho(\mathbf{r}_2) 
        \left[1- C \exp({-\gamma \left|\mathbf{r}_1-\mathbf{r}_2\right|}) \right]
        }
                     {\left|\mathbf{r}_1-\mathbf{r}_2\right|} d\mathbf{r}_1 d\mathbf{r}_2
%        \textcolor{red}{w(\eta,r12)} 
        \label{eq:ecoul-yuk}
        \end{equation}
and 
        \begin{equation}
        E^{\text{Yuk},\gamma}_x  
        = -\frac{1}{2} \iint 
               \frac{
        \rho(\mathbf{r}_1) \rho_x(\mathbf{r}_1,\mathbf{r}_2) 
        \left[1- C \exp({-\gamma \left|\mathbf{r}_1-\mathbf{r}_2\right|}) \right]
        }
                     {\left|\mathbf{r}_1-\mathbf{r}_2\right|} d\mathbf{r}_1 d\mathbf{r}_2
%        \textcolor{red}{w(\eta,r12)}    
        \label{eq:eexch-yuk}
        \end{equation}
where $C$ is a constant. 
Using these, the correlation energy $E_c$ is then given by 
%by introducing a Yukawa type Coulomb hole in the Hartree and the
%        exchange-energy expressions,
        \begin{align}
        E_{c} 
	& = 
%        \left. 
        (E^{\text{Yuk},\gamma}_H + E^{\text{Yuk},\gamma}_x)
%\right|_{\text{Coulomb hole}} 
	- 
        (E^{\text{Yuk},\gamma=0}_H + E^{\text{Yuk},\gamma=0}_x)
%        (E^{HS}_{\textrm{H}} + E^{HS}_x)
	\\
        & = - \frac{C}{2} \iint 
               \frac{
        \rho(\mathbf{r}_1) 
%        \textcolor{red}{
        \left[
        \rho(\mathbf{r}_2) + \rho_x(\mathbf{r}_1,\mathbf{r}_2)
        \right]
        \exp({-\gamma \left|\mathbf{r}_1-\mathbf{r}_2\right|})
        }
%        }
                     {\left|\mathbf{r}_1-\mathbf{r}_2\right|} d\mathbf{r}_1 d\mathbf{r}_2
%        \textcolor{red}{w(\eta,r12)}    
        \nonumber	\\
	&= C \bar{E}_{\text{corr}}
	\label{eq:constC}
        \end{align}
%where $E^{HS}_{\textrm{H}}$ is the Hartree energy in the Harbola-Sahni approach and $E^{HS}_x$ is the corresponding 
%exchange energy.
Comparing the above equation with the~\eref{eq:ecdft}, we have for the Coulomb correlation hole 
\begin{align}
        \rho_{c}(\mathbf{r}_1,\mathbf{r}_2) = 
        \rho_{c}(\gamma,\mathbf{r}_{12}) = 
        - C 
	\left[
	\rho(\mathbf{r}_2) + \rho_x(\mathbf{r}_1,\mathbf{r}_2)
	\right]
        \exp({-\gamma \left|\mathbf{r}_1-\mathbf{r}_2\right|})
\end{align}
Similar to the Chakravorty and Clementi Coulomb hole, 
the above correlation hole also doesn't satisfy the charge neutrality condition~(\eref{eq:rhocorr-3}). 
In addition, 
%exact constraints but 
the above Coulomb hole does not go to zero in the limit $\gamma \rightarrow 0$. 
 
%Moreover, the 
%$E_H^{Yuk}$ and $E_x^{Yuk}$ have a repulsive term in the limit $r_{12} \rightarrow 0$, 
%whereas in Chakravorty and Clementi's work, the above terms goes to zero. 

In the following, we proposed a model form for Coulomb correlation hole which goes to zero as required.
Furthermore, it is also has a term so that it satisfies the charge neutrality condition. 
The proposed model Coulomb correlation hole
        \begin{align}
        \rho_c(\gamma,r_{12}) = 
        \rho_c(\mathbf{r}_1,\mathbf{r}_2) = 
	C 
        \left[
        - \rho(\mathbf{r}_2) + \rho_x(\mathbf{r}_1,\mathbf{r}_2)
        \right]
\exp({-\gamma \left|\mathbf{r}-\mathbf{r}'\right|}) \sin(2\gamma \left|\mathbf{r}_1-\mathbf{r}_2\right|) 
	\label{eq:coulombhole-mod}
        \end{align}
which 
%inaddition to satisfying the constraints, all 
goes to zero in the limit $\gamma \rightarrow 0$. 
The factor $\sin(2\gamma \left|\mathbf{r}_1-\mathbf{r}_2\right|)$ is 
reminiscent of Friedel oscillations near a defect in a solid~\cite{book:ziman:1972}. 
%, and the corresponding Hartree and exchange has a proper form. 

In our calculations, the parameter $\gamma$ in the model is to be tuned to satisfy the charge neutrality.
        \begin{equation}        
        \int \rho_c(\mathbf{r}_1,\mathbf{r}_2) d\mathbf{r}_2 = 0 \, \, \text{for all} \, \, \mathbf{r}_1
	\label{eq:rhoc-cond1}
        \end{equation}
In an inhomogeneous system, we replace condition~(\eref{eq:rhoc-cond1}) by 
%        Making an alternative condition which makes independent of $\mathbf{r}_1$,
        \begin{equation}        
        \iint \rho_c(\mathbf{r}_1,\mathbf{r}_2) d\mathbf{r}_1 d\mathbf{r}_2 = 0
        \end{equation}
which makes it independent of $\hemvec{r}_1$. 
        The parameter $\gamma$ in the Coulomb correlation hole is now chosen to satisfy this
        condition.
In the following, we first apply our method to ground-states to check its validity. 
We then extend it to excited-states to explore its applicability there. 

\section{Ground-state results}
\begin{figure}
%       \begin{subfigure}[b]{1.00\textwidth}
                \centering
%%              \scalebox{1.6}{
%               \input{Figures/pl-ex-rdfvx-Li-3s1-2S.tex}
%%               } 
%               \resizebox{1.0\textwidth}{!}{
%%%%            \resizebox{!}{.5\paperheight}{
%       \input{Figures/ke-hlike-gr.tex}
%%        \xput[0.5]{
%                \resizebox{!}{.35\paperheight}{
                \input{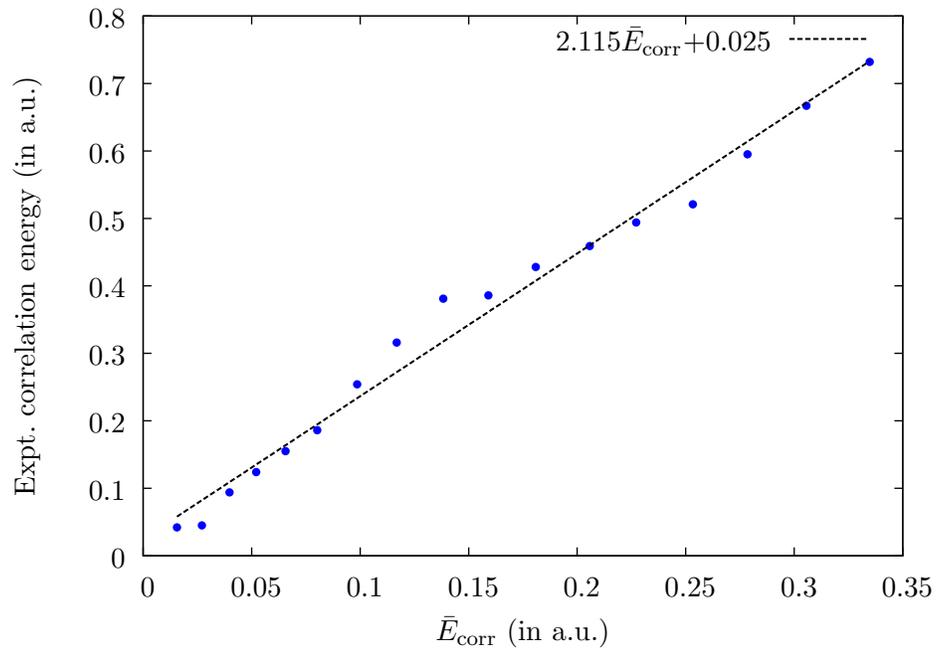}
%                }
%%        }
        \caption
%	[Optional caption ]
	{
		Plot of calculated $\bar{E}_{\text{corr}}$ and the experimental correlation energies. 
		The dotted line is the linear fit of the data. 
        }
        \label{fig:corr-fit}
%               \input{\figip/gr_vx_beta_0pt073/pl-gr-vx-Li.tex}
%%%%            }
%               \end{subfigure}%
\end{figure}

We now use the correlation hole of~\eref{eq:coulombhole-mod} to 
calculate the correlation energies. 
For this, the orbitals obtained from the Harbola-Sahni exchange-only calculations are used. 
Shown in~\tref{tab:corr-gr} are the results 
%optimized parameter $\gamma$  
obtained by tuning the parameter $\gamma$ 
in the modelled correlation hole of~\eref{eq:coulombhole-mod} to satisfy the charge neutrality constraint. 
$\bar{E}_{\textrm{corr}}$ obtained from~\eref{eq:eexch-yuk} corresponding to the optimized $\gamma$ 
are also shown in the table. 
% are the experimental correlation energies. 
The unknown normalization factor in the modelled Coulomb hole 
is obtained by taking the ratio of the $\bar{E}_{\textrm{corr}}$ and the 
experimental correlation energies. 
%obtained using the 
It is worth noting that factor $\text{Expt.}/\bar{E}_{\textrm{corr}}$
%in the model Coulomb correlation hole
is nearly independent of $Z$ 
and is maximum for Li from an average value close to $2.3$.  
This is also evident from~\fref{fig:corr-fit} where the 
experimental correlation energies and the $\bar{E}_{\textrm{corr}}$ are plotted. 
The dotted line is the linear fit of the data, with the slope equal to $2.115$. 
%$2.11533*x+0.0248777$. 
% ratio
%and are compared with the experimental correlation energies with the
%-------------------------------------------------
%       f(x)=2.11533*x+0.0248777 \approx 2.115*x+0.025
%\multicolumn{6}{X}{NEUTRAL ATOMS}      \\
%-------------------------------------------------
\begin{table}
% ----------------- TABLE ----------------- 
	\centering
	\caption
%	[corr-gr]
	{
		Correlation energies of atoms in their ground-states.  
		Numbers given are in atomic units.
%		$f(x)=2.11533*x+0.0248777 \approx 2.115*x+0.025$
	}
%\begin{threeparttable}
%\begin{ThreePartTable}
       \begin{tabular}{|p{3cm}|p{2cm}|p{3cm}|p{3cm}|p{3cm}|}
%       \begin{tabular}{|l|r|r|r|r|}
%	\begin{tabu} to \textwidth {X[l]HHHX[r]X[r]cc}
%	\begin{tabu} to \textwidth {X[l]HHHX[r]X[r]XX}
\hline
		Atom
		&$\gamma$
		&-$\bar{E}_{\textrm{corr}}$
		&-Expt.
		&Expt/$\bar{E}_{\textrm{corr}}$  \\
\hline
He      &5.2    &0.0156 &0.042  &2.69   \\
Li      &8.0    &0.0271 &0.045  &1.67   \\
Be      &10.8   &0.0398 &0.094  &2.36   \\
B       &13.6   &0.0521 &0.124  &2.38   \\
C       &16.3   &0.0656 &0.155  &2.36   \\
N       &18.9   &0.0802 &0.186  &2.32   \\
O       &21.2   &0.0986 &0.254  &2.58   \\
F       &23.6   &0.1168 &0.316  &2.70   \\
Ne      &25.8   &0.1383 &0.381  &2.82   \\
Na      &28.2   &0.1591 &0.386  &2.43   \\
Mg      &30.6   &0.1809 &0.428  &2.36   \\
Al      &32.8   &0.2058 &0.459  &2.23   \\
Si      &35.3   &0.2272 &0.494  &2.17   \\
P       &37.5   &0.2533 &0.521  &2.06   \\
S       &39.8   &0.2785 &0.595  &2.14   \\
Cl      &42.0   &0.3056 &0.667  &2.18   \\
Ar      &44.1   &0.3348 &0.732  &2.36   \\
\hline
%	\end{tabu}
       \end{tabular}
%	\begin{tablenotes}
%		\footnotesize
	%	\item[a] \label{tn:p86-corr} PRB.33.8822 (1986) \\ J.Phys.B:At.Mol.Opt.Phys. 20 3599 (1987)
	%	\item[a] \label{tn:lyp-corr} LYP - PRB.37.785 (1988)
%		\item[a] \label{tn:expt-corr-neutral} For neutral atoms, J.Phys.B:At.Mol.Opt.Phys. 20 3599 (1987).
%		\item[b] $\gamma/Z \approx 2.6$ for all the atoms considered. 
	%	\item[c] \label{tn:expt-corr-ions} LYP - PRB.37.785 (1988)
%	\end{tablenotes}
	\label{tab:corr-gr}
%\end{threeparttable}
%\end{ThreePartTable}
\end{table}

In the following section, we use this scaling factor to estimate the 
correlation energies of atoms in their excited-states. 
%Using the ground-state, we further went and calculated the correlation enegies for excited-states.

\clearpage
\subsection{Results for excited-state correlation energies}
Similar to the ground-state calculations, the orbitals obtained 
from the Harbola-Sahni potential are used to calculate the correlation energies 
for excited-states. 
%$\bar{E}_{\text{corr}}$. 
Shown in~\tref{tab:corr-ex-en} are the results obtained for excited-states of atoms 
by tuning the parameter $\gamma$ in the modelled Coulomb hole to satisfy the exact constraint. 
The correlation energies obtained using the ground-state LYP functional are also shown in the table. 
Also shown in the last column is the correlation energies obtained from~\eref{eq:ecorr-def} using the 
Harbola-Sahni and the Hartree-Fock exchange-energy respectively. 
%re calculated using 
The exact non-relativistic energies are taken from the Monte-Carlo calculations~\cite{galvez-buendia-sarsa:2002,galvez-buendia-sarsa:2005}. 

\begin{table}[h]
% ----------------- TABLE ----------------- 
	\centering
	\caption
%	[corr-ex]
	{
		Correlation energies of atoms in their 
		excited-states. 
		Numbers given are in atomic units.
	}
%\begin{threeparttable}
%\begin{ThreePartTable}

%       \begin{tabular}{p{3cm}p{2cm}p{3cm}p{3cm}}
       \begin{tabular}{|p{3cm}|p{2cm}|p{2cm}|p{2cm}|p{2cm}|p{2cm}|p{2cm}|}
%       \begin{tabular}{lrrrrrr}
%	\begin{tabu} to \textwidth {X[l]X[r]ccccc}
%	\begin{tabu} to \textwidth {X[l]X[r]X[r]X[r]X[r]X[r]X[r]}
	\hline
		Atom
		&$\gamma$
		&-$\bar{E}_{\textrm{corr}}$
%		&Present work 
		&-$2.115\bar{E}_{\textrm{corr}}$
		&-$E_c^{\text{LYP}}$
		&\multicolumn{2}{c}{-$E_c$}	\\
%                \cmidrule(l{05pt}r{01pt}){6-7}        %l{5pt}r{10pt}
%                \cmidrule(l{25pt}r{01pt}){6-7}        %l{5pt}r{10pt}
		&&&&
                &HS%~\tnote{a}
                &HF%~\tnote{b}	
\\
%		\midrule
		%-------------------------------------------------
% EXCITED  STATES
%-------------------------------------------------
%Atom				&$\gamma$&Ebar_corr	&Ebar_corr*2.115	&LYP		&Sarsa-HS	&Sarsa-HF	\\
%\midrule                                                                                                       
Li$(2p^1 \ ^2P)$		&8.0    &0.0265		&0.05605		&0.05338	&-		&-		\\      
C$(2s^1 2p^3 \ ^3D)$		&16.1   &0.0650		&0.13747		&0.16181	&-		&-		\\      
N$(2s^1 2p^4 \ ^4P)$		&18.6   &0.0801		&0.16941		&0.19305	&-		&-		\\      
O$(2s^1 2p^5 \ ^3P)$		&21.0   &0.0971		&0.20537		&0.26547	&-		&-		\\      
F$(2s^1 2p^6 \ ^2S)$		&23.2   &0.1170		&0.24745		&0.32642	&-		&-		\\      
Ne$^+(2s^1 2p^6 \ ^2S)$		&25.6   &0.1359		&0.28743		&0.34319	&-		&-		\\      
Li$(3s^1 \ ^2S)$		&8.0    &0.0268		&0.05668		&0.04986	&-		&-		\\      
Li$(4s^1 \ ^2S)$		&8.0    &0.0267		&0.05647		&0.04856	&-		&-		\\      
B$(3s^1 \ ^2S)$			&13.7   &0.0522		&0.11040		&0.11102        &-		&-		\\      
%\midrule                                        		                                
B$(3p^1 \ ^2P)$			&13.7   &0.0520		&0.10998		&0.10994	&-		&-		\\      
%\midrule                           	                                                
B$(2s^1 2p^2 \ ^2D)$		&13.5   &0.0512		&0.10829		&0.12873	&0.12778	&0.12398	\\      
C$^+(2s^1 2p^2 \ ^2D)$		&16.2   &0.0645		&0.13642		&0.14479	&0.12936	&0.12586	\\          
N$^{2+}(2s^1 2p^2 \ ^2D)$	&18.9   &0.0782		&0.16539		&0.15563	&0.13176	&0.12843	\\         
O$^{3+}(2s^1 2p^2 \ ^2D)$	&21.6   &0.0919		&0.19437		&0.16353	&0.13260	&0.12941	\\          
F$^{4+}(2s^1 2p^2 \ ^2D)$	&24.3  &0.1059		&0.22398		&0.16958	&0.13173	&0.12864	\\          
Ne$^{5+}(2s^1 2p^2 \ ^2D)$	&27.0   &0.1199	        &0.25359		&0.17438	&0.12898	&0.12599 	\\   
%\midrule                                   		                                
Be$(2s^1 2p^1 \ ^3P)$		&10.8   &0.0387		&0.08185		&0.06538	&0.05867	&0.05574	\\
Be$(2s^1 3s^1 \ ^3S)$		&10.9   &0.0385		&0.08143		&0.06227	&0.05429	&0.05257	\\
Be$(2s^1 3p^1 \ ^3P)$		&10.9   &0.0387		&0.08185		&0.06232	&0.05284	&0.05010	\\
Be$(2s^1 3d^1 \ ^3D)$		&10.9	&0.0383		&0.08100		&0.06161	&0.05259	&0.05048	\\

%		\bottomrule
%		\bottomrule
	\hline
%	\end{tabu}
       \end{tabular}
%	\begin{tablenotes}
%		\footnotesize
%		\item[\textdagger] $2.115*\bar{E}_{\textrm{corr}}$
%                \item[a] HS : $E_{exact}-E_{x}^{HS}$; 
%		$E_{exact}$ from~\cite{galvez-buendia-sarsa:2002,galvez-buendia-sarsa:2005}
%		\item[b] HF : $E_{exact}-E_{x}^{HF}$; 
%                \item[a] HS : $E^{Sarsa}_{exact}-E_{x}^{HS}$; HF : $E^{Sarsa}_{exact}-E_{x}^{HF}$
%		\item[b] \label{tn:aa} 
%		$E_{exact}$ from~\cite{galvez-buendia-sarsa:2002,galvez-buendia-sarsa:2005}
		%\item[b] Another note
%	\end{tablenotes}
	\label{tab:corr-ex-en}
%\end{threeparttable}
%\end{ThreePartTable}
\end{table}

The $\gamma$ is observed to be almost the  same for a given atomic number  and is state-independent. 
For example, $\gamma$ is equal to $8.0$ for all the excited-states of Li, for Boron, out of four 
excited-states considered, $\gamma$ is $13.5$ for one case and is equal to $13.7$ for the rest of the three cases.
However, applying it further to estimate the correlation energies of excited-state atoms are
not quite accurate. 
A further study is required.
One reason for this, is the ground- and excited-state correlation energies are almost similar.
\section{Concluding remarks}
In this chapter, we have tried to estimate the 
correlation energies of various atoms in their excited-states. 
For this, the Coulomb hole is modelled in terms
of the orbitals following the previous work by Chakravorty and Clementi.
The parameter in the model is fixed by making
the corresponding Coulomb hole satisfy the
exact constraint of charge neutrality.

The ground-state results obtained with this modelled Coulomb hole 
is shown to be indenpendent of $Z$. 
Extending the ground-state parameter to the excited-states, 
we have calculated the excited-state correlation energies. 
The correlation energies so obtained for excited-states 
in majority of cases match with the exact values. 
Only for ions with high ionicity they do not match with the 
exact values. 
% in most of the cases is different 
%from the exact correlation energies. 
%did not lead to 
A further study is required. 

Other systematic approach to calculate the correlation energies is through the response function calculations. 
We plan to take this approach in the near future for estimating the correlation energies of excited-states. 
%our attempt

%We then compare the 
%It is shown that the correlation energies obtained
%by us are similar to those given by the LYP functional.
\bibliographystyle{plain} 
%\bibliographystyle{natbib} 

%\bibliography{mkh.bib,final.bib} 

\end{document}